\documentclass{aa}
\usepackage{graphicx}
\usepackage{amsmath}
\usepackage[normalem]{ulem}
\begin{document}
   \title{The radio galaxy $K$-$z$~relation: the 10$^{12}\,\mathrm{M_\odot}$\ 
mass limit}
\subtitle{Masses of galaxies from the $\mathrm{L_K}$\ luminosity, up to $z >$4} 
\authorrunning{B. Rocca-Volmerange et al.}
\titlerunning{10$^{12}\,\mathrm{M_\odot}$~Galaxies and $K-z$~ Relation}
   \author{B. Rocca-Volmerange
          \inst{1,3}
          \and
          D. Le Borgne \inst{1}
	\and
	C. De Breuck \inst{1}
	\and
	M. Fioc \inst{1}
       \and
         E. Moy\inst{2}
}
   \offprints{brigitte.rocca@iap.fr}
\institute{Institut d'Astrophysique de Paris,
	98bis Bd Arago, 75014 Paris, France
         \and
             DSM/DAPNIA, Service d'Astrophysique, CEA-Saclay, B\^at 709, 
91191 Gif-sur-Yvette, France
	\and Universit\'e de Paris-Sud XI, B\^at 333, 91405 Orsay cedex, France
	   }

\date{Received 28 April 2003; accepted 7 November 2003}

\abstract{ The narrow $K$-$\mathrm{z}$\ relation of powerful radio galaxies
in the  Hubble $K$\  diagram is often attributed to  the stellar
populations  of  massive elliptical  galaxies. Because  it extends 
over a  large range of redshifts  (0 $< z
<$4), it is difficult to
estimate masses  at high redshifts  by taking into account galaxy
evolution.  In the present paper,  we propose to estimate the stellar
masses  of  galaxies using the galaxy  evolution  model
P\'EGASE. We use star  formation scenarios that successfully fit
faint  galaxy counts  as well  as  $z$=0 galaxy  templates. These
scenarios  also
predict  spectra at  higher $z$,  used to  estimate  valid photometric
redshifts.    The   baryonic   mass   of  the   initial   gas   cloud
$M_{\mathrm{bar}}$\  is then  derived.  The
$K$-$\mathrm{z}$\  relation is remarkably  reproduced by  our 
evolutionary scenario
for elliptical galaxies  of baryonic mass $M_{\mathrm{bar,max}} \simeq
$10$^{12}\,\mathrm{M_\odot}$, at   all   $  z$~up   to   $z$=
4. $M_{\mathrm{bar,max}}$ is also the maximum mass limit of all types
of  galaxies.   Using another  initial  mass  function  (IMF), even  a
top-heavy  one, does  not alter  our  conclusions. The  high value  of
$M_{\mathrm{bar,max}}$\  observed at  $z  >$ 4  implies that  massive
clouds  were already formed  at early  epochs. We  also find  that the
$M_{\mathrm{bar,max}}$\  limit   is  similar  to   the  critical  mass
$M_{\mathrm{crit}}$\ of a  self-gravitating cloud regulated by cooling
(Rees  \& Ostriker,  1977; Silk,  1977).  Moreover,  the  critical size
$r_{\mathrm{crit}}\simeq $  75 Kpc is remarkably close  to the typical
diameter of Ly$\alpha$~haloes surrounding distant radio galaxies. This   confirms  the  validity   of  the   method  of   baryonic  mass
determination based on the  $K$-band luminosity.  A puzzling question
that remains to be answered is the short time-scale of mass-accumulation required to
form  such massive  galaxies at  $\mathrm{z}=$  4. We discuss  the
dispersion of the  $K$-$z$\ relation in terms of  uncertainties on the
mass limit. The link between the  presence of the active nucleus and a
large   stellar   mass   is  also   discussed.
     }

\maketitle

\begin{figure*}    
\centering    
\includegraphics[width=8.5cm]{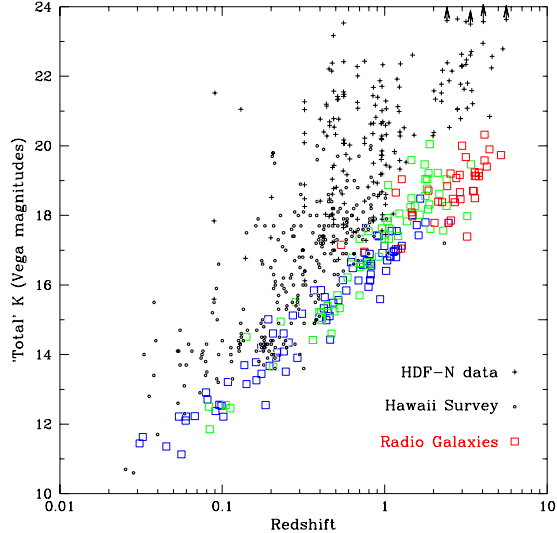}
\caption{Radio galaxy hosts (empty squares) and deep field galaxies 
(dots and crosses) plotted in the Hubble $K$\ diagram. The catalogues of 
radio galaxies are respectively 3CR (blue squares) and 6CE (green squares).
The sample also includes highly distant radiogalaxies observed with the 
NIRC and some 6C$^*$ radio galaxies (red squares). 
The field galaxy samples are the HDF-N (black crosses, Williams et al., 1996) and Hawaii (black dots, Cowie et al., 1999) surveys.}
\label{figure:Kz_sansmodel}
\end{figure*}

\section{Introduction}
The $K$\ Hubble diagram is an efficient tool
to study the evolution of galaxies at high redshifts. Because 
evolutionary processes in the expanding universe
are not known, models are required. The most valuable models propose scenarios of star formation that aim to fit observations of galaxies at all redshifts; the first constraint is to reproduce
the templates of nearby ($z \simeq$0) galaxies. 

The 
$K$-band (centred at 2.2 $\mu$m) is preferred for evolutionary analysis because galaxies are seen over
a large redshift range (0$\leq z \leq $ 4) in their near-IR to optical rest frames.  
Moreover, the evolutionary effects are reduced. 
The Hubble $K$-$\mathrm{z}$ diagram is known to be an  excellent tool to  
measure stellar masses of galaxies, up to high redshifts.  
At $\mathrm{z \simeq}$0, the $K$\ luminosities are  dominated  by  
the bulk of old low mass red stars. At higher 
$\mathrm{z}$, the $K$\ emission is due to the redshifted emission of blue
luminous young stars because galaxies are more gas-rich and formed stars more 
actively in the past. 
 
The main  feature of  the galaxy distribution  in the  Hubble $K$-band
diagram is the well-known $K$-$\mathrm{z}$\ relation, the sharp bright
limit traced by nearby ($z\simeq$  0) massive galaxies and by hosts of
distant  powerful  radio  galaxies.    The  physical  meaning  of  the
relation,  up to  $z $=  4 or  more, is  still puzzling.  Most  of the
interpretations of the $K$-$\mathrm{z}$\  relation explain its
dispersion either  by testing various sets  of cosmological parameters
(H$_0$,  q$_0$,  $\Lambda_0$)  (Longair  \&  Lilly, 1984,  and  more
recently Inskip et al. 2002)  or by examining the relation of 
the galaxy $K$\  magnitude  with its radio  power.  The  radio-powerful  3CR
galaxies are  on average brighter ($\Delta  K \simeq -$  0.6 mag) than
the  less radio-powerful  6C (Eales  \& Rawlings,  1996; Eales  et al,
1997) or 7C (Willott  et al. 2003) radio galaxies.  Moreover the 
radio-powerful Molonglo sample  (McCarthy, 1999)  shows a  $\Delta K
\simeq -$  1 mag with  the 6C catalogue  at all redshifts. From
their  near-infrared morphologies, the  stellar populations  of radio
galaxy hosts are identified  with massive elliptical galaxies, even at
high  redshifts  (van  Breugel  et  al.,  1998;  Lacy  et  al.,  2000;
Pentericci  et al.,  2001).  Radio galaxy  luminosities are  generally
high, typically  L $\simeq  $ 3  to 5 L$_*$\  for both  radio-loud and
radio-quiet  galaxy  hosts (Papovich  et  al.,  2001). Radio
galaxy  hosts are  located  in  place of  elliptical  galaxies in  the
fundamental plane  (Kukula et al.,  2002). At high  $\mathrm{z}$, mass
estimates of radio galaxy  hosts are rare.  From  an incomplete
rotation   curve,   Dey   et   al.,   (1996)   derived   a   mass   of
10$^{11}\,\mathrm{M_\odot}$  for  3C  265  ($\mathrm{z}$=0.81).  Using
HST/NICMOS observations, Zirm et al., (2003) confirmed that radio galaxy
hosts  are massive,  even at  high $\mathrm{z}$. Stellar masses
estimated by fitting stellar energy distributions (SEDs) to the NICMOS
data are  between 3 and  8 10$^{11}\,\mathrm{M_\odot}$. The  CO (3-2)
line of the very massive  galaxy SMM J02399-0136 (z=2.8) observed with
the IRAM  observatory by  Genzel et  al., (2003) shows  that it  is a
rapidly rotating  disk with  a total dynamical  mass of $\simeq$  3 ×
10$^{11}$    sin$^2$    i    M$_\odot$    ($i$\    is    the    galaxy
inclination). Stellar masses of three  galaxies, selected in
the  Hubble Deep  Field South  at  K$_s <$~22  on the  basis of  their
unusually red  near-IR color (J-K $>$  3), were estimated  to be about
10$^{11}$ M$_\odot$~from their SEDs (Saracco et al., 2003).

Section  2  presents the  observational  samples  of  radio and  field
galaxies  in  the  Hubble   $K$\  diagram.   Following  a  preliminary
interpretation  of the  $K$-$\mathrm{z}$\  relation (Rocca-Volmerange,
2001),  we explore  in section  3 the  space of  cosmology  and galaxy
evolution  parameters   with  the   code  P\'EGASE  to   confirm  this
interpretation.  We  successively analyse the  parameter set: distance
modulus,  star  formation  scenarios that define  the  evolutionary
time-scales  of  the  various  components,  the IMF  and  finally  the
baryonic  mass of  the initial  gas reservoir. For  the  sake of
clarity,  evolution scenarios  are limited  to spiral  and elliptical
types: considering scenarios between  these two types would not modify
our conclusions. In  section 4, the $K$-$\mathrm{z}$\ diagram is
analysed with the predicted  $K$-$\mathrm{z}$\ sequences of spiral and
elliptical  galaxies formed  from clouds  of various  baryonic masses
$M_{\mathrm{bar}}$=$10^{9}$\   to~10$^{12}$\,$\mathrm{M_\odot}$.    We
find that only  the sequence of the elliptical  galaxies formed from a
10$^{12}$\,$\mathrm{M_\odot}$\  progenitor cloud  uniformly  fits the
$K$-$\mathrm{z}$\  relation from  $z$=0 to  $z$=4. The  sensitivity of
this  result  to  other  factors  (cosmology  models,  intensities  of
emission lines) is considered.
 
The           striking           similarity           of           the
 $M_{\mathrm{bar,max}}$=10$^{12}$\,$\mathrm{M_\odot}$\ mass limit to
 the theoretical  estimate of the  critical mass of  fragmentation for
 self-gravitating clouds (Rees \& Ostriker, 1977) is finally discussed
 in section 5.  Since  the $K$-$\mathrm{z}$\ relation is mainly traced
 by the most powerful radio galaxies, this means that the most massive
 galaxies  correspond to  the  most massive  black  holes.  Masses  of
 galaxies of various radio powers are derived in section 6. In particular,
 strong constraints on mass accumulation time-scales are given by the high
 masses of  radio galaxy  hosts at $\mathrm{z}  \simeq$4.  The discussion
 and        conclusion make up   the        last       sections.
   \begin{figure*}
\includegraphics[width=6cm,angle=90]{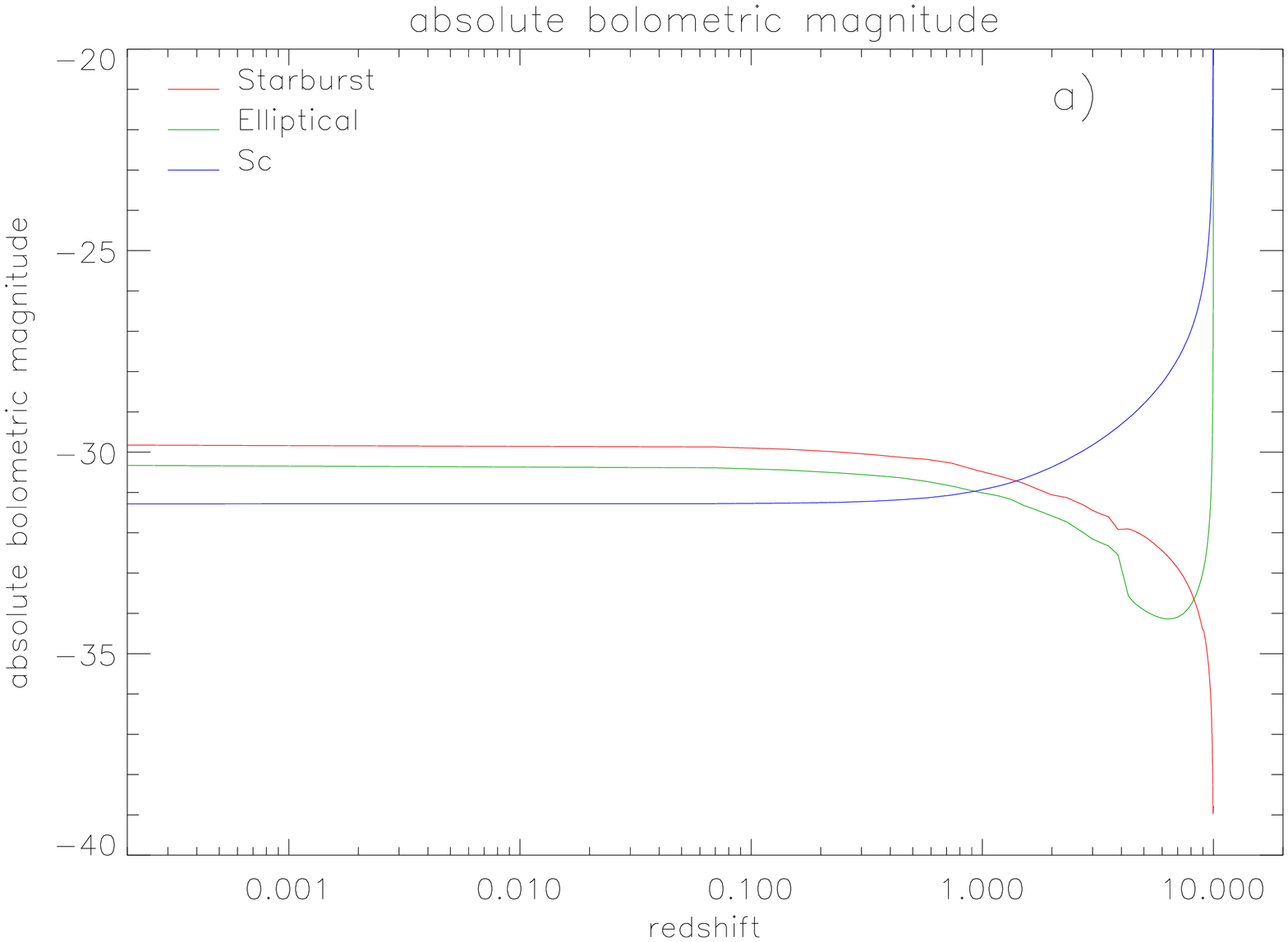}
\includegraphics[width=6cm,angle=90]{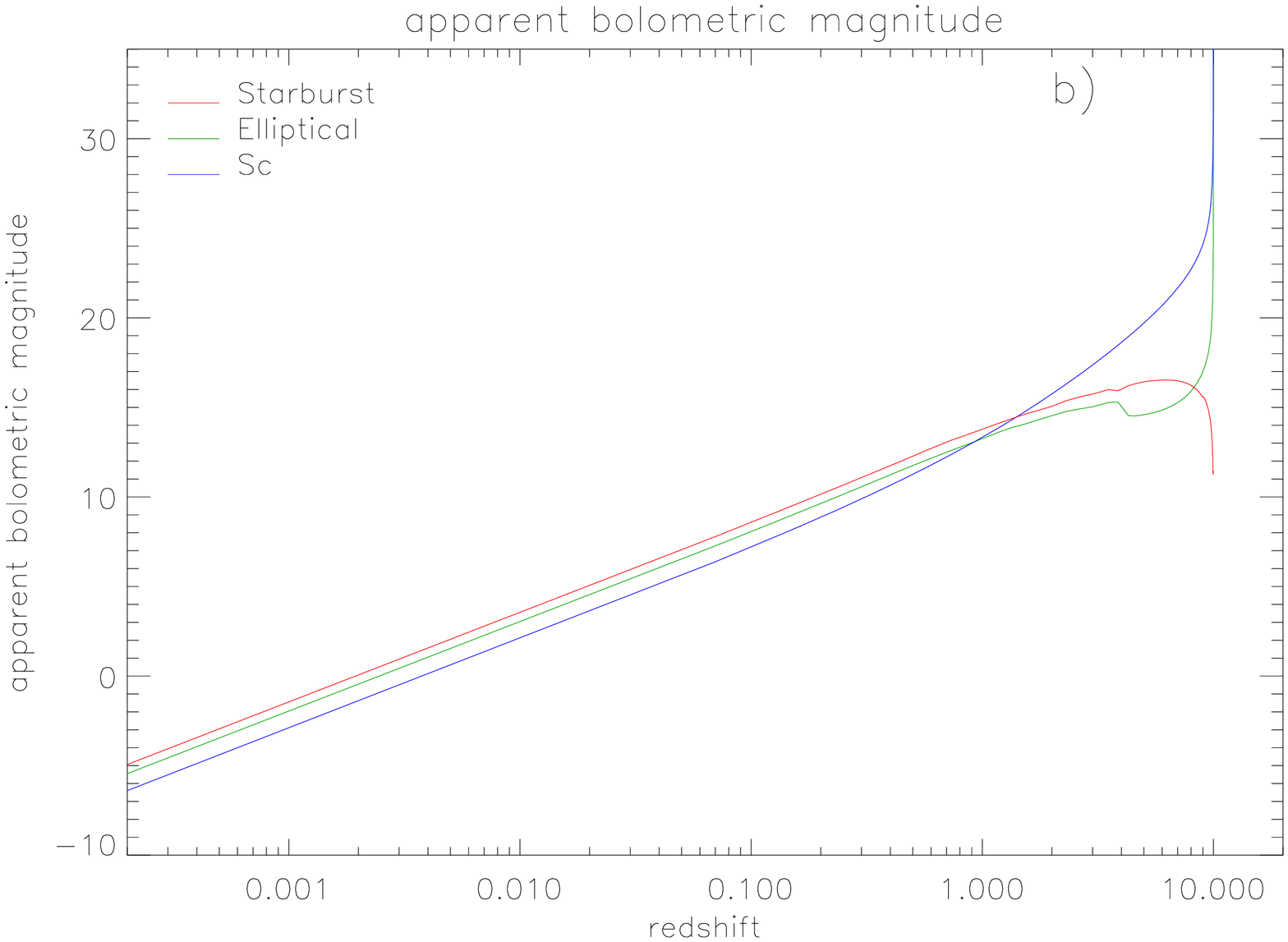}
   \caption{(a) Plot of  the absolute    bolometric 
magnitude $\mathrm{M_{bol}(z,t(z))}$\ vs  $\mathrm{z}$\
for starburst, spiral Sc and elliptical template galaxies.
(b) similar plot with the apparent  bolometric magnitude
$\mathrm{m_{bol}(z,t(z))}$. The comparison of the two plots shows that
the distance modulus is the primary explanation for 
the $K$-$z$~slope up to $\mathrm{z \simeq}$\ 1, 
whatever the spectral type. At higher $\mathrm{z}$, stellar evolution 
has a major effect on light emission for all spectral types.}
\label{figure:DModul}
\end{figure*}
 
\section{The observational $K$-$\mathrm{z}$~relation}
 
The observational  sample of distant field galaxies  and of radio-loud
 galaxies  in the  $K$-band has  been gathered  by De  Breuck  et al.,
 (2002) (Fig. \ref{figure:Kz_sansmodel}).  Radio  sources are from the
 3CR and 6CE catalogues completed by radio galaxies observed with NIRC
 at  the Keck  I telescope  (Laing et al.,  1983, Eales  et al.,
 1997, van Breugel et al.,  1998). All the magnitudes of radio sources
 are corrected  to a  standard 64 kpc  metric aperture,  following the
 method of Eales et al., (1997). The brightest ($z\simeq $1) 3CR
 sample is complete. The NIRC sample  is a selection by van Breugel et
 al., (1998) of  powerful radio sources at the  highest redshifts (1.8
 $< z <$ 4.4) with strong emission lines.  The 6CE and NIRC catalogues
 are  not complete.   In sections  4  and after,  the complete  7C
 catalogues and the 6CE* sample, recently published by Willott et al.,
 (2003), were added to the De Breuck's sample for the interpretation of
 the  $K$-$z$\  relation.  Over  the  8-magnitude  range of  $K$-band
 luminosities  of galaxies, the  bright envelope  of the  radio galaxy
 distribution  delineates a  sharp  limit, up  to $\mathrm{z}$=4:  the
 so-called $K$-$\mathrm{z}$~relation.   For comparison, field galaxies
 of  the Hawaii  survey  (Cowie et  al.,  1999) and  the HDF-N  survey
 (Williams  et  al.,  1996),  observed  in the  $K$-band  at  KPNO  by
 Dickinson et  al., (2003)  are also plotted  in the  Hubble diagram.
 The  spectroscopic redshifts  were published  by  Fern\'andez-Soto et
 al., (1999), Cohen et al., (2000) and Songaila et al., (1994).

 \section{Modeling   galaxy  evolution   through  the   $K$-band}  
The
variation of  apparent $K$\ magnitudes  with redshift is  predicted by
galaxy evolution  models, assuming  a cosmological model.   To clarify
the sensitivity  of the  model outputs to  the adopted  star formation
scenarios,  we  separately analyze  the  effects  of other  parameters
(distance  modulus,  IMF,  initial  mass  and  redshift,  cosmological
parameters).  Stellar  energy distributions (SED)  of galaxy templates
are predicted at all redshifts. The main emission lines, due to
ionization by massive stars, are  computed with an HII region model in
which the Lyman  continuum photon number and metallicity  are the main
parameters. Their intensities  are added to the SEDs,  assuming a FWHM
of 10 \AA\ at $z$=0.

For each spectral type, the reddened apparent magnitude of a synthetic
galaxy through  the filter ${\lambda}$\ at redshift  ${z}$\ and cosmic
time $t(z)$\ is given by:
\[
\begin{split}
m_{\lambda}(z,t(z))=&M_{\lambda}(0, t(0))\\
&+(m-M)_\mathrm{bol}(z)+A_{\lambda}(z) + k_{\lambda}(z) +e_{\lambda}(z)
\end{split}
\]
where M$_{\lambda}$(0, t(0)) is  the intrinsic magnitude at $z=0$\ and
at   the   present   cosmic   time   t(0).    The   distance   modulus
$(m-M)_{bol}$(z)   will   be   considered  below.   The   k-correction
$k_{\lambda}(z)$\  and   e-correction  $e_{\lambda}(z)$\  respectively
depend on the cosmology and on the evolution scenario.  The extinction
A$_{\lambda}(z)$\ through the filter  ${\lambda}$\ is derived from the
total amount of metal  and  dust  geometry. In  our  code P\'EGASE,  the
extinction factor for elliptical galaxies is computed with a radiative
transfer code for  a dust distribution fitted on  a King's profile and
Draine's  grain model (see  Fioc, 1997 for details).  For spirals,
the stars, gas and dust are mixed homogeneously.

 We adopt  the cosmological model  of Friedmann-Lema\^\i tre  with the
Hubble  constant H$_0$ in  km.  s$^{-1}$.   Mpc$^{-1}$ and  the matter
$\Omega_{M}$\  and cosmological  constant  $\Omega_{\Lambda}$\ density
parameters.   We   successively  consider  the   H$_0$,  $\Omega_{M}$,
$\Omega_{\Lambda}$\ values:  (65, 0.3, 0.7)  for standard ,  (65, 0.1,
0.)  for  open and (65,  1., 0.) for  flat universes.  Note  that 
evolutionary models are able to successfully reproduce a wide range of
galaxy observables only in  the open and non-null $\Omega_{\Lambda}$\
universes.

\subsection{The distance modulus}
The  distance  modulus $\mathrm{(m-M)_{bol}}(z)$\  is  related to  the
luminosity  distance  $D_\mathrm{L}(z)$,   and  depends  only  on  the
cosmological  parameters (Weinberg, 1972).   Fig.  \ref{figure:DModul}
shows  the  absolute (a)  to  apparent  (b)  magnitude change  in  the
standard  cosmology for three  spectral types  (starburst, elliptical,
spiral Sc). At low $z$, the evolutionary  effects of early-type
galaxies are negligible because the  emission is dominated by the bulk
of low-mass giant stars. The slight luminosity increase for spirals is
due to recent star formation. At higher $z$, evolution effects become
dominant  when star formation  rates are  more intense,  in particular
when  $ \mathrm{z}\simeq  \mathrm{z_{for}}$  (=10 in  our model);  the
galaxy  formation redshift  $\mathrm{z_{for}}$\ is  the  redshift when
simultaneously infall and initial star formation begin.

\begin{figure*}    
\centering    
\includegraphics[width=10cm]{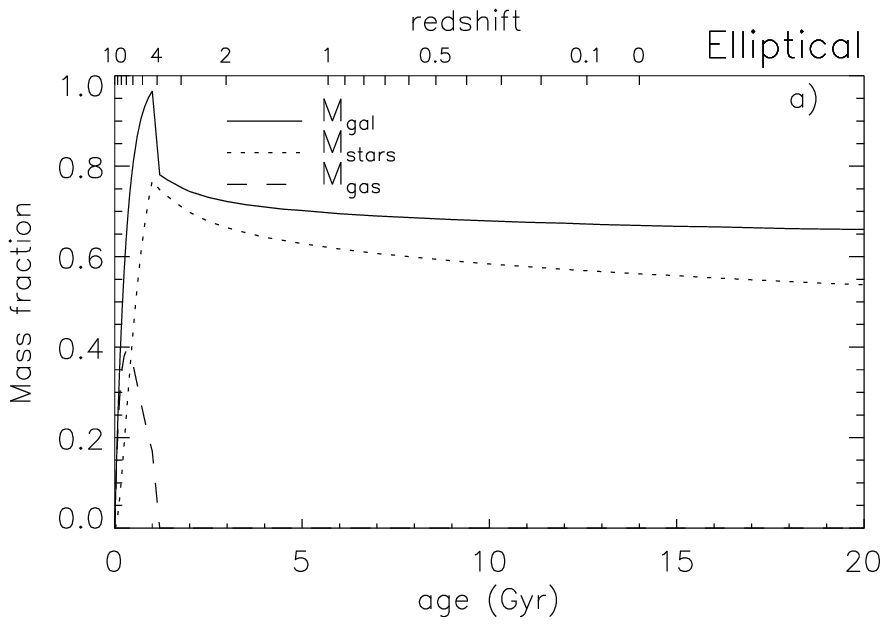}
\includegraphics[width=10cm]{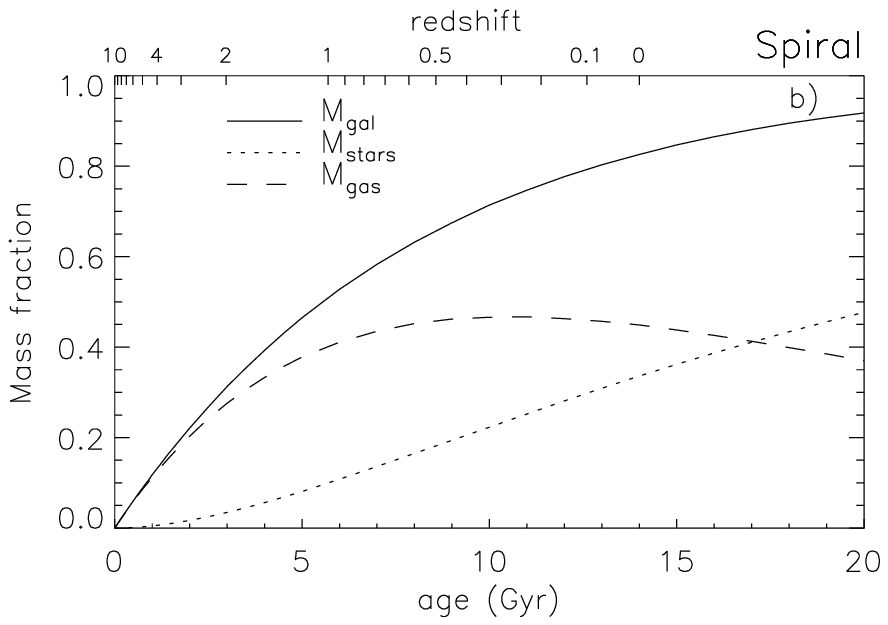}
\caption{Evolution of mass fractions ($M_{\mathrm {stars}}$,
$M_{\mathrm{gas}}$ and 
$M_{\mathrm{gal}}$=$M_{\mathrm{stars}}$+$M_{\mathrm{gas}}$+
$M_{\mathrm{remnants}}$) for elliptical (a) and spiral (b) galaxies. 
Ages in Gyr and corresponding redshifts for a standard cosmology are
indicated. The formation redshift is $\mathrm{z_{for}}$= 10. 
All masses vary between 0 to 1, after dividing by 
the initial gas cloud of mass $M_{\mathrm{bar}}$. 
Peaks observed at about 1 Gyr in the elliptical plots result
from the stopping of star formation activity by galactic winds following
the infall phase.  This scenario fits the template SED 
of ${\mathrm{z \simeq}}$ 0 elliptical galaxies (see text for details).}
\label{figure:Masses}
\end{figure*}
\subsection{Star formation rates and mass fractions}
In the  following, $M_\mathrm{bar}$\  is the initial  mass of  the gas
reservoir, the  baryonic mass of the progenitor  cloud.  The scenarios
of   star  formation  conservatively   assume  star   formation  rates
proportional to  the current  gas mass. The duration of star
formation  decreases   from  irregular  and  spiral   ($>10$  Gyr)  to
elliptical ($<1$Gyr  ) galaxies. The classical formalism  is only 
briefly reviewed here.  The number of stars formed per log
mass unit and time unit is
\[
\mathrm{d}^2N(m,t) = \mathrm{SFR}(t) \times \mathrm{IMF}(m) \mathrm{d}\log(m)\mathrm{d}t.
\]
$\mathrm{SFR(t)}$  =   $M_\mathrm{gas}(t)$/$\tau_{sf}$\  is  the  star
formation law of the  galaxy where $M_\mathrm{gas}(t)$\ is the current
gas  mass  of  the  galaxy  and $\tau_{sf}$\  is  the  star  formation
efficiency. IMF  is the  initial mass function  and is assumed  not to
evolve. All variables  are normalized to the mass  unit of the initial
cloud,  so  that  the  model outputs  are  multispectral  luminosities
$\mathrm{(L/M)_{\lambda}}$\     and    mass     fractions    (stellar:
$M_\mathrm{stars}$,  gaseous:  $M_\mathrm{gas}$).   More  details  are
given in  the README  of the code  P\'EGASE.2 \footnote  {Available at
http://www.iap.fr/pegase}  (Fioc  \&  Rocca-Volmerange, 1997,  1999c).
For simplicity, among the  eight scenarios of galaxy
evolution proposed by P\'EGASE.2, we only select those for elliptical
and spiral Sc galaxies.   The extreme difference between their star
formation efficiencies  $\tau_{sf}$\ implies different  star formation
durations  ($\simeq$~1Gyr   for  ellipticals   and  $>$  10   Gyr  for
spirals). This is long enough to constrain the accumulation
time-scales of  the stellar and  progenitor masses. Another  reason to
adopt the  two scenarios is  that the formation  law of our  spiral Sc
scenario  is  close  to  the  star  formation  law  deduced  from  the
hierarchical  model  by  Baron  \&  White (1987)  while  our  law  for
ellipticals  is drastically  different  to hierarchical  predictions.
Note that the instantaneous  starburst, often adopted to model stellar
populations of radio galaxy hosts  (Ridgway et al., 2001, Willott
et al., 2003), is unphysical.  Moreover, because the time-scale is too
short, the  scenario gives  too red colors  when they are  compared to
those of massive elliptical galaxies, typical of powerful radio galaxy
hosts.

Fig. \ref{figure:Masses} presents, respectively for elliptical (a) and
spiral (b)  galaxies, the  evolution with age  and $z$\ of  the galaxy
mass  $M_\mathrm{gal}$\  and of  its  stellar $M_\mathrm{stars}$\  and
gaseous $M_\mathrm{gas}$~masses.  For the two types, star formation 
starts at infall from the initial gas reservoir. It simultaneously
initiates  galaxy   formation,  so   that  the  initial   galaxy  mass
$M_\mathrm{gal}$\  is null.  $M_\mathrm{gas}$\  measures the  gas mass
within the galaxy. Neutral atomic  HI and molecular H$_2$ hydrogen are
not   distinguished.  The   stellar   mass  $M_\mathrm{stars}$\   only
corresponds  to stars  still alive  while stellar  remnant  masses are
taken into  account in $M_{\mathrm{remnants}}$.   $M_\mathrm{gal}$\ is
the    sum    of    $M_\mathrm{stars}$\    +    $M_\mathrm{gas}$\    +
$M_\mathrm{remnants}$.     Template     elliptical    scenarios    are
characterized by  a decreasing infall time  scale of 300  Myr and
the star  formation efficiency $\tau_{sf}$=0.3  $Myr. M_\odot^{-1}$.
Then  supernovae produce  strong  galactic winds  starting  at 1  Gyr,
eliminating  all gaseous components  and instantaneously  stopping any
further star  formation. Note that our  scenarios do  not assume
further    inflow    of   cool    gas    from    the   halo.     From
Fig.   \ref{figure:Masses}a),   the   elliptical  scenario   is   then
characterized by  peaks of  $M_{gal}$\ and $M_{stars}$\  as $M_{gas}$\
drops  because  galactic winds. After  the  age of  1Gyr,  the
difference between  $M_{gal}$\ and $M_{stars}$\ of  ellipticals is due
to      the     integrated      mass      of     stellar      remnants
$M_{\mathrm{remnants}}$.  In  template Sc  spirals,  the infall  time
scale is 8 Gyr and the efficiency is $\tau_{sf}$= 10$Myr. M_\odot
^{-1}$\ without  galactic winds.  The star formation  time duration is
much longer  ($>$ 10 Gyr). The baryonic mass  M$_{bar}$\ of the
progenitor  is  derived  by  fitting the  apparent  $K$\  luminosities
$\mathrm{L_{K}}$\     of     galaxies     with    the     predictions
$\mathrm{(L/M)_{K}}  \times M_{bar}$\  for $10^{9}  \leq  M_{bar}/\mathrm{M_\odot} \leq
10^{13}$.

\subsection{Initial mass function}
The     IMF    parameters    are     the    slope     parameter    $x$
($d\mathrm{N(m)}/dlog\mathrm{m}\  \alpha\   \mathrm{m}^{x}$)  and  the
lower and  upper mass limits.  Three IMFs are considered.   First, for
Rana   \&    Basu's   (1992) IMF,   the   slope    parameter   above   1.4
$\,\mathrm{M_\odot}$~is    $x$=-1.51   and    $x$=-1.71    above   6.5
$\,\mathrm{M_\odot}$. Second, for  Salpeter's (1959) IMF,
the slope parameter is $x$= -1.35.  Finally, for Gibson \& Matteucci's
(1997) IMF,  $x$=  -0.8,  which  greatly  increases  the  mass  of  metal
ejected. This "top-heavy"  IMF was proposed by the  authors to explain
high metal  abundances at  the most remote  epochs.  For all  IMFs, we
adopt a lower  mass of 0.1 $\,\mathrm{M_\odot}$\ and  an upper mass of
120$\,\mathrm{M_\odot}$.
  
A truncated IMF, biased towards massive  stars, with a lower mass of 1
$\,\mathrm{M_\odot}$\  and  a  slope   of  -0.8  (the  upper  mass  is
unchanged) is  unable to reproduce  the bright $K$-band  magnitudes of
nearby elliptical galaxies: the bulk of low mass stars that dominates
the stellar emission at $z$=0 in the $K$-band is then missing. It will
not be considered below.
 

\section{The $\mathrm{K}$-$\mathrm{z}$~relation} 
\begin{figure*}    
\centering   
\includegraphics[width=8.5cm]{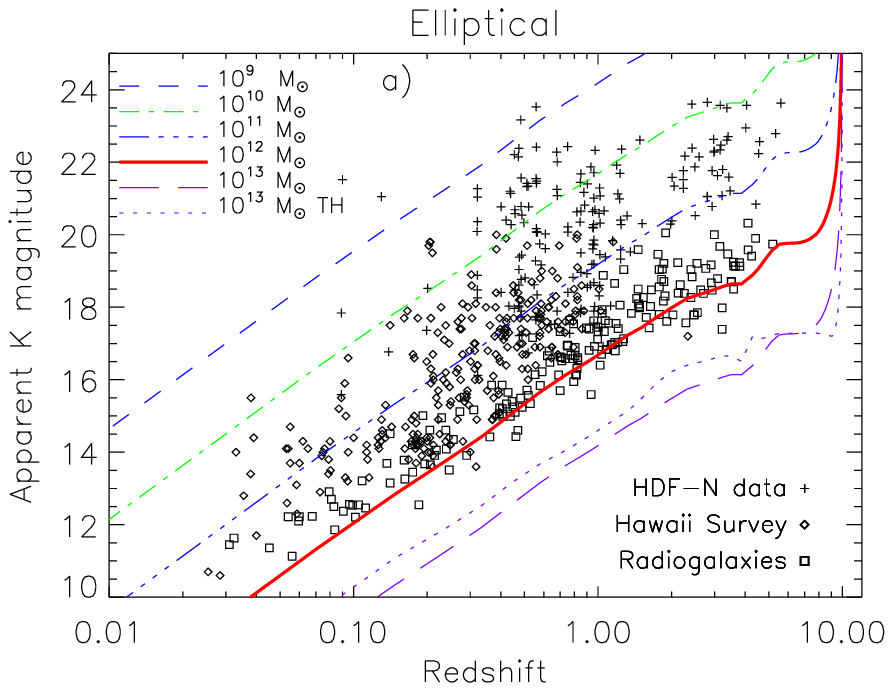}
\includegraphics[width=8.5cm]{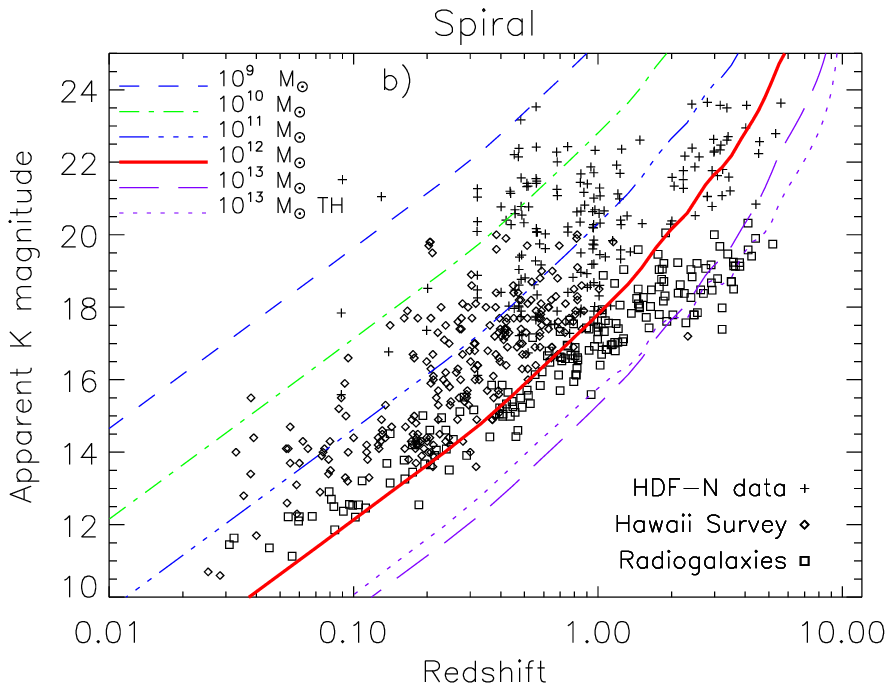}
\caption{$K$-$\mathrm{z}$~sequences for  
$M_{\mathrm{bar}}$ = 10$^{9}$~to 10$^{13}\,\mathrm{M_\odot}$\ compared
to field and radio galaxy samples in the Hubble $K$-band\ diagram.  The IMF
is from Rana \& Basu, 1992, except for $M_{\mathrm{bar}}$ = 10$^{13}\,\mathrm{M_\odot}$,
the top-heavy IMF is also used (dotted line).
The 10$^{12}\,\mathrm{M_\odot}$\
elliptical sequence (a) coincides with the $K$-$\mathrm{z}$\
 limit, up to $\mathrm{z} >$ 4.
The spiral galaxy scenarios (b) are unable to
fit distant radio galaxies at high $\mathrm{z}$\ except for 
$M_{\mathrm{bar}}$ = 10$^{13}\,\mathrm{M_\odot}$, this value is however
unacceptable at low $\mathrm{z}$. 
}
\label{figure:Kz_Mass}
\end{figure*}

Figure  \ref{figure:Kz_Mass} compares  the  observed distributions  of
galaxies in the Hubble $K$\ diagram with the predictions of elliptical
(a)  and spiral  (b) models.  The  model sequences  are computed  with
constant  values  of M$_{\mathrm  {bar}}$\  in  the range  10$^{9}$~to
10$^{13}$\,$M_{\odot}$ for the two scenarios.  The IMF is from Rana \&
Basu     (1992)    for    all     sequences,    except     for    the
10$^{13}$\,$\mathrm{M_\odot}$\   sequence  also   computed   with  the
"top-heavy" (TH) IMF.

All observations are covered by the sequences
of    $M_{\mathrm   {bar}}$    between    10$^{9}$\   and    10$^{12}\
\mathrm{M_\odot}$; so the  initial  mass  of  the  reservoir  cloud
M$_{\mathrm  {bar}}$\  is  the   main  parameter explaining  the
evolution of stellar masses in the Hubble $K$-band diagram.

Fig. \ref{figure:Kz_Mass}a also shows the continuous fit of the bright
$K$-$z$\  relation   by  the   sequence  of  elliptical   galaxies  of
$M_{\mathrm{bar,max}}\simeq$    10$^{12}\,\mathrm{M_\odot}$,   up   to
$\mathrm{z}>$\   4.  For  spirals,   the  10$^{12}\,\mathrm{M_\odot}$\
sequence (Fig.  \ref{figure:Kz_Mass}b) is  not acceptable at  high $z$
while   the  $10^{13}\,\mathrm{M_\odot}$\   one,   possible  at   high
$\mathrm{z}$, has no match at low $\mathrm{z}$.

Confirming  that  powerful  radio   galaxies  are  hosted  by  massive
ellipticals (van Breugel et al. 1998; De Breuck et al. 2002; Willott
et    al. 2003), we    also   estimate    their   stellar
$M_{\mathrm{stars}}$\  and galaxy  $M_{\mathrm{gal}}$\  masses at  all
redshifts:     for     ellipticals      at     $z$=0     (14     Gyrs)
(Fig.   \ref{figure:Masses}a)  $M_{\mathrm{stars}}\simeq$\   55\%  and
$M_{\mathrm{gal}}\simeq$\ 65\% of the initial mass $M_{\mathrm{bar}}$\
of  the gaseous progenitor; these values  only vary  within $\pm$10\%
between  $z$=4 and  $z$=0.   Thus $\simeq$\  35\% of  the
baryonic    mass     is    still     in    the    halo     and   
$M_{\mathrm{remnants}}\simeq$\  10\%. Gas  exchanges as  late inflows
may occasionally change these  fractions.  However, depending on their
efficiency,  they  would induce  fluctuations  of  the star  formation
history incompatible  with the  uniformity of  the $K$-$z$\
relation.  

The scenario  of spirals  (Fig. \ref{figure:Masses}b) gives,  at $z$=0
(14 Gyrs), $M_{\mathrm{stars}}\simeq$\ 35\%, $M_{\mathrm{gas}}\simeq$\
45\%   and  $M_{\mathrm{gal}}\simeq$\   85\%  of   the   initial  mass
$M_{\mathrm{bar}}$.   These masses rapidly  evolve with  redshift: at
$z\simeq$2, $M_{\mathrm{stars}}\simeq$\ 3\%, $M_{\mathrm{gas}}\simeq$\
27\%   and  $M_{\mathrm{gal}}\simeq$\   31\%  of   the   initial  mass
$M_{\mathrm{bar}}$.

Finally,  field galaxies  of deep  surveys are  fitted either  by less
massive  ``elliptical''  scenarios  of  $M_{\mathrm{bar}}$=  10$^{9.5\
{\mathrm{to}}\ 11}\,\mathrm{M_\odot}$\  or by ``spiral''  scenarios of
$M_{\mathrm{bar}}$=10$^{10\ {\mathrm{to}}\ 12}\,\mathrm{M_\odot}$. The
Hubble $K$\  diagram is alone unable  to separate the  two models.  In
particular at low  $z$ ($<$ 0.3), the predictions  of the most massive
elliptical and spiral models are similar.

\begin{figure*}    
\centering   
\includegraphics[width=8.5cm]{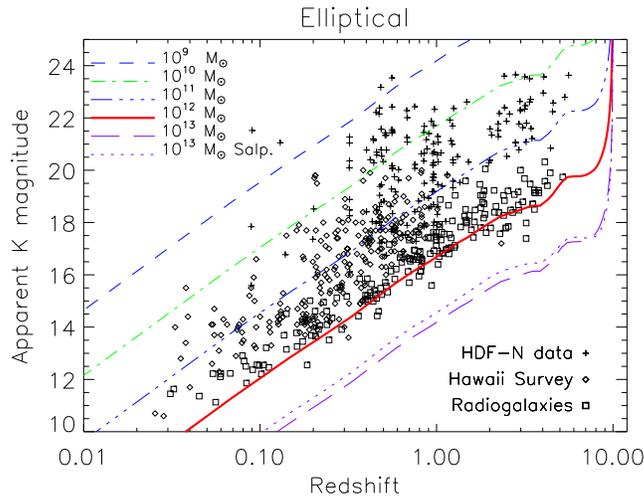}
\caption{The $M_{\mathrm{bar}}$= 10$^{13}\,\mathrm{M_\odot}$\ sequence
shows the comparison of Rana \& Basu IMF(long-dashed line) with Salpeter IMF(dotted line).
The rest is similar to Fig. \ref{figure:Kz_Mass}a. 
} 
\label{figure:Kz_Salp}
\end{figure*}

The  $K$-$z$\  sequences  are  predicted  with two  other  IMFs.   The
variation  of results computed with either
Rana  \&  Basu's IMF or  the  "top-heavy" (TH) IMF  is
presented  on  the  M$_{\mathrm{bar}}$ =  10$^{13}\,\mathrm{M_\odot}$\
sequence  in  Fig.  \ref{figure:Kz_Mass}ab. Fig.  \ref{figure:Kz_Salp}
also  shows   the  M$_{\mathrm{bar}}$  =  10$^{13}\,\mathrm{M_\odot}$\
sequence with  the Salpeter IMF compared to  Rana \& Basu.  The IMF
effect on the $K$\ luminosity is less than 10\% at all ${z}$.

\begin{figure*}    
\centering   
\includegraphics[width=16cm]{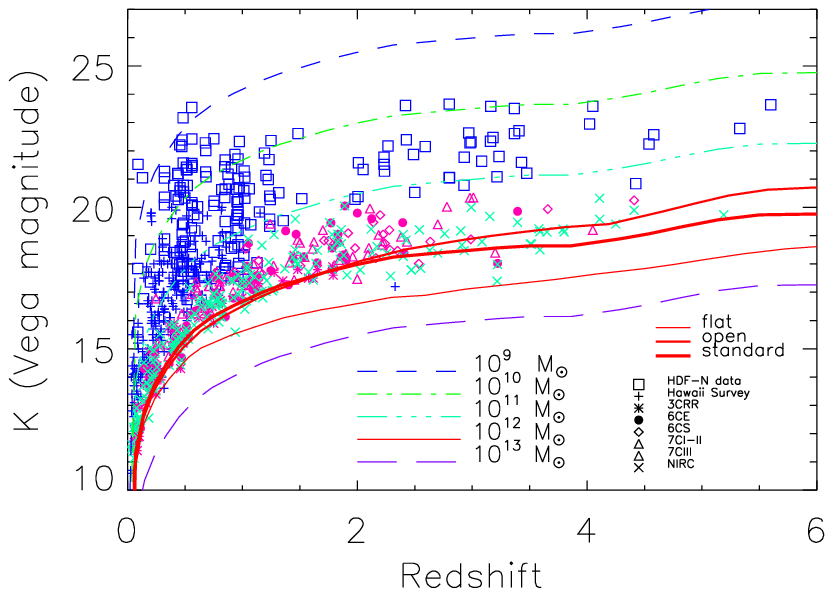}
\caption{Interpretation of the 
$K$-$z$~diagram with models of various baryonic masses as in 
Fig. \ref{figure:Kz_Mass} or Fig \ref{figure:Kz_Salp} but $z$\ is 
in a linear scale. The catalogues 7C-I and 7C-II (Willott et al. 2003), 7C-III (Lacy et al. 2000)
and 6C$^*$~(Jarvis et al. 2001), are added  to the sample from  
De Breuck et al., (2002). 
The sensitivity to  
cosmological parameters is analyzed on the  
10$^{12}\,\mathrm{M_\odot}$~sequence 
of elliptical galaxies (red full lines) for 3 cosmologies (open, flat 
and standard). Only open and standard (flat with a cosmological constant) universes are acceptable by data.  
}
\label{figure:Kz_cosmo}
\end{figure*} 

 
The  $M_{\mathrm{bar,max}}\ $sequence  limit  for massive  ellipticals
depends little on the formation redshift $z_{\mathrm{for}}$.  Changing
$z_{\mathrm{for}}$\ ($\geq $6.6, see Hu  et al., 2002; Kodaira et al.,
2003) in elliptical scenarios  modifies the $K$-$\mathrm{z}$\ relation
only during the first Gyr (see Fig. \ref{figure:Kz_Mass}a).

The   standard   cosmology   (H$_0$=65   km.   s$^{-1}$.   Mpc$^{-1}$,
$\Omega_{M}$=0.3, $\Omega_{\Lambda}$=0.7)  has been adopted.  However,
to study  the sensitivity  to the cosmological  model, we  compare the
$K$-band observations to evolution models  in the $z$ (and not log $z$
as  previously shown) scale  (Fig. \ref{figure:Kz_cosmo})  for various
universes.   A flat universe  without the cosmological  constant (H$_0$=65
km. s$^{-1}$. Mpc$^{-1}$, $\Omega_{M}$=1, $\Omega_{\Lambda}$=0) is not
compatible with the data.   Both standard and open universes can fit
the data;  the variation of $M_{\mathrm{bar,max}}$\ given  by the two
models is lower than 10 \%.

Fig. \ref{figure:Kz_cosmo}  is similar to  Fig. \ref{figure:Kz_Mass}ab
and Fig.  \ref{figure:Kz_Salp} but in the redshift  scale and includes
more  data: the  catalogues 7C-I  and  7C-II (Willott  et al. 2003),
7C-III (Lacy et al. 2000)  and 6C$^*$~(Jarvis et al. 2001) are added
to  the  De Breuck  compilation.   All  our  previous conclusions,  in
particular  that the  more powerful  radio galaxy  hosts are  the more
massive galaxies, up  to 10$^{12}\,\mathrm{M_\odot}$, are confirmed by
these data.  

\begin{figure*}    
\centering   
\includegraphics[width=9.5cm]{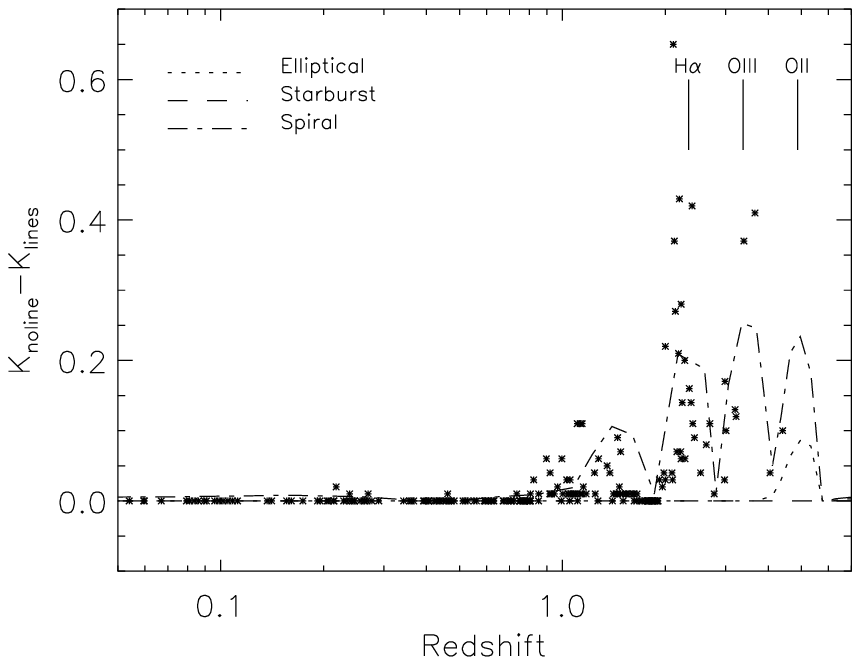}
\caption{Predictions through the $K$-band of the main emission lines 
H$\alpha$,  [OIII]$\lambda$5007,
[OII]$\lambda$3727 
from photoionization by massive stars
for spirals and 
ellipticals. Comparison data are the 3CR, 6C and 7C  
catalogues of radio sources (Willott et al. 2003). 
Galaxy ages depend on redshifts defined by lines
($z$=2.3, 3.3 and 5 for H$_{\alpha}$, [OIII] and [OII]). 
The excess between observations 
and models is due to other ionization sources (shocks, AGN).
} 
\label{figure:emissionlines}
\end{figure*}

Emission lines from gas  photoionization by massive stars are computed
in our models.   Fig. \ref{figure:emissionlines} shows predictions for
the    main    emission    lines   (H$\alpha$,    [OIII]$\lambda$5007,
[OII]$\lambda$3727), redshifted  in the  $K$-band and compared  to the
data.       The     $K$-band     luminosity      variation     $\Delta
K$=$K_{noline}-K_{lines}$,  due to ionization  by massive  stars, does
not  exceed   0.25  mag  in  spirals,  as   in  ellipticals.  However,
observations  show  stronger emission  lines  than model  predictions:
other sources  of ionization (shocks  and photoionization by  AGN) are
known to be highly efficient in radio galaxies. From the powerful NIRC
and  the  7C  radio  sources  (Fig.  \ref{figure:emissionlines}),  the
$\Delta K/K$ relative excesses due  to emission lines are $\leq$5\% at
high $z$. 


\section{The $K$-$z$\ relation and the fragmentation limit}
The     uniform    galaxy    mass     limit    $M_{\mathrm{bar,max}}$=
10$^{12}\mathrm{M_\odot}$\   at  all   redshifts  from   the  $K$-band
luminosities  is striking. The  critical mass  of fragmentation  for a
self-gravitating  cloud  has  been  estimated  by  Rees  \&  Ostriker
(1977). The dynamical  model predicts that the kinetic  energy of infall
is  first thermalized via  shocks; further  evolution then 
depends on the relative cooling $t_{cool}$~  and free-fall $t_{ff}$~
time-scales.    High    masses   in    the    range   10$^{10}$\    to
10$^{12}\mathrm{M_\odot}$\ ,undergoing an  efficient  cooling,
collapse at about the free-fall rate, fragment and possibly form stars
while larger  masses may experience a  quasi-static contraction phase;
they go into free fall only when they reach critical values of radius
and mass. The authors evaluated the critical mass $M_{crit}\ \simeq$ \
10$^{12}\mathrm{M_\odot}$ between  the two regimes  that define the
fragmentation limit.They  also estimated the  mass-independent critical
size $r_{crit}\ \simeq$75 Kpc.

The  maximum baryonic  mass  $M_{\mathrm{bar,max}}$\ delimitating  the
observed cut off  of the galaxy distribution in  the Hubble diagram is
then similar to the theoretical fragmentation limit of galaxies.  This
clarifying   explanation   of    the   $K$-$z$\   relation   is   very
simple.  Moreover the  corresponding critical  size $r_{\mathrm{crit}}
\simeq   $  75   Kpc   remarkably  fits   the   typical  diameter   of
Ly$\alpha$~haloes  (150 Kpc) surrounding  active radio  galaxies.  The
value  $M_{\mathrm{bar,max}}$\ at  $z  \simeq $0  is also  dynamically
confirmed by measurements of masses of nearby galaxies.

This correspondance between the various mass estimates 
emphasizes the robustness of
mass  determination using $K$-band stellar luminosities.
This  method of  mass  determination is thus
potentially  appropriate to  all deep  galaxy surveys which 
may improve our knowledge of mass evolution at high $z$.
 
The $M_{\mathrm{bar,max}}$\ limit for radio galaxies is also estimated
from  observations,  up  to  $z  >$  4  (Fig.   \ref{figure:Kz_Mass}),
implying  strong  constraints on  galaxy  evolution  models. The  most
massive galaxies  which reach the  fragmentation limit at  $z \simeq$4
formed their stellar  mass within 1Gyr.  This time-scale  is too short
to  form the  stellar  mass  by the  hierarchical  merging of  initial
$\simeq 10^{5}\mathrm{M_\odot}$\  building blocks as  predicted by the
CDM  model.  However these  objects might  be rare  and their  mode of
formation  exceptional.   On the  other  hand,  because distant  radio
galaxies are  known to  be in  a dense medium  of already  evolved and
luminous (masses $\simeq 10^{11}\mathrm{M_\odot}$?) galaxy companions,
the  self-gravitational  model  may  efficiently accumulate
several 10$^{11}\mathrm{M_\odot}$\ within 1 Gyr if the cooling process
is highly efficient: the  dissipation by strong emission lines of
the huge  (diameter of $\simeq$ 150  Kpc) ionized cocoon  of the radio
galaxies  may  explain  the  cooling efficiency  at  high  redshifts.
Chokshi (1997) already evoked  the Rees \& Ostriker (1977) models
to interpret  distant radio galaxies,  relating them to  large density
fluctuations undergoing isothermal free-fall  and regulated by
cooling H and He  line emission. Her interpretation suggested intense
stellar formation triggered by the  radio jet. The present analysis of
the  Hubble $K$-diagram that confirms  large stellar  populations in
distant  radio galaxies is  unable  to affirm  that massive  star
formation was  triggered by the radio  jet, in the  AGN environment or
outwards  from  the  overpressured  cocoon.   Only high  spatial
resolution on extended  radio galaxies will help to clarify such
hypotheses.
  
\section{Masses of radio galaxy hosts}
\begin{figure*}    
\centering   
\includegraphics[width=9.5cm]{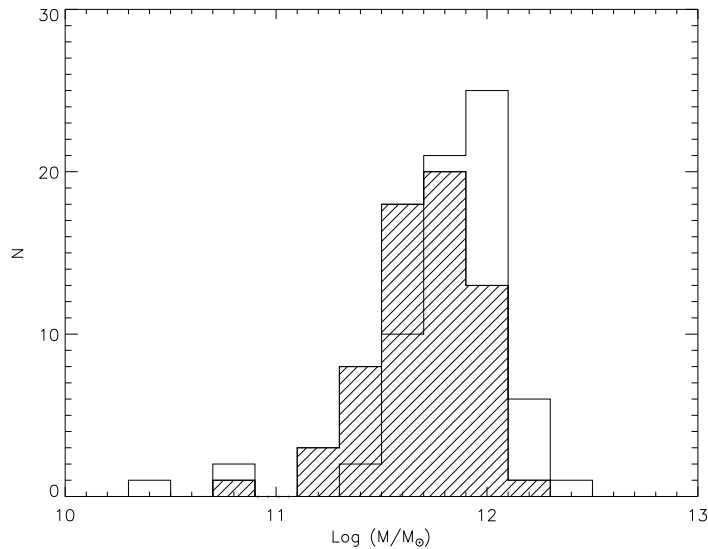}
\caption{Baryonic masses of 3C and 7C galaxies estimated
from the elliptical scenario. The more radio-powerful 3CR galaxies 
(empty histogram) 
{\bf are on} average more massive than the less radio-powerful 7C galaxies
(grey line histogram). The difference 
$\mathrm{D [Log (M_{\mathrm{bar}}}$/ $\mathrm{M_\odot})]
\simeq$ 0.2 corresponds to $\Delta M_{\mathrm{bar}} \simeq$\ 2 10$^{11}\mathrm{M_\odot}$.
} 
\label{figure:Kz_Willott}
\end{figure*}

In  the literature,  the  stellar  masses of  radio  galaxy hosts  are
tentatively related to their radio  power: as an example, spheroids of
stellar  masses greater than  4  10$^{11}\mathrm{M_\odot}$\ are
more frequently  observed in  radio-loud than in  radio-quiet galaxies
(McLure  \&  Dunlop, 2002). Willott  et  al. (2003)  found
systematic differences between the $K$-band magnitude distributions in
the  Hubble  diagram  of  the  7C,  6CE, 6C*  and  3CR  catalogues  of
increasing radio-power.  The 7C distribution is 0.55  mag fainter than
the  3CR  at  all  redshifts.  Note that  such  a  $K$-band  magnitude
difference  at all redshifts  means in  our interpretation  an average
mass     difference    between     the     two    catalogues.      Fig
\ref{figure:Kz_Willott} shows the respective mass distributions of the
3CR catalogue  (empty histogram)  and of the  7C catalogue  (grey line
histogram) computed  with our  models.   The  two distributions  are
limited    by     the    mass    limit    $M_{\mathrm{bar,max}}\simeq$
10$^{12}\,\mathrm{M_\odot}$, confirming that  the limit is independent
of radio power.   The more radio-powerful 3CR catalogue  is on average
more massive  (11.7 $<$ Log  ($M_{\mathrm{bar}} / \mathrm{M_\odot}$)$<
$12.1)   than  the   less  powerful   7C  catalogue   (11.5   $<$  Log
($M_{\mathrm{bar}}  /  \mathrm{M_\odot}$)$<$  12.0).   The  respective
peaks  of the  two distributions  differ by  ~0.2, corresponding  to a
variation      of     $\Delta      M_{\mathrm{bar}}\simeq      $     2
10$^{11}\,\mathrm{M_\odot}$.   The   variation  is  not  significantly
sensitive to cosmological or IMF effects.

The  stellar and  baryonic mass  estimates from  $K$-band luminosities
with  the  help  of  evolution  models is  limited  by  uncertainties.
According  to  Fig.  \ref{figure:Kz_cosmo},  the  predicted  sequences
computed  with the  open and  standard universes  are  compatible with
observations within  10\% at the highest $z$.  As previously analyzed,
the stellar  masses of ellipticals  vary within 10\% depending  on the
adopted IMF. Another source of uncertainty at high redshifts is 
the  calibration  at $z$=0.   Depending  on templates  $M_{\lambda}(0,
t(0))$ that are  either the  brightest cluster  members  or averaged
elliptical galaxies,  the uncertainty may reach  10\%. Strong emission
lines at precise redshifts are  seen in the Hubble $K$~diagram so that
they  require a  special  fitting procedure.   More  uncertain is  the
nebular continuum  of radio  sources observed in  the blue  band (rest
frame), not modelled in elliptical galaxies. On the basis of starburst
models, it may be lower than 20\% at $z>$4.  The global uncertainty on
the upper limit of  baryonic mass of  galaxies given by  models is
$\simeq$  50\%.  That  implies  $M_{\mathrm{bar,max}}$=  1  $\pm$  0.5
10$^{12}\,\mathrm{M_\odot} $. For less luminous galaxies, the accuracy
will depend on evolutionary scenarios by types.

\section{Discussion}
The main results  obtained in this analysis are  the following: first,
the   $K$-$z$~relation    is   due   to   the    galaxy   mass   limit
10$^{12}\,\mathrm{M_\odot}$;  second,   fitting  the  $K$-$z$~sequence
corresponds  to  the fragmentation  limit  predicted by  gravitational
models;  finally,  we  are   able  to  identify  massive  galaxies  of
10$^{12}\,\mathrm{M_\odot}$~ at redshifts  $z\simeq$4.  The mass limit
from Rees  \& Ostriker, (1977) based on  dynamical arguments validates
the baryonic mass determination based on the $K$-band luminosity.
\begin{itemize}
\item
Estimates are obtained using star formation scenarios with infall
from the baryonic halo,  winds, active and passive stellar evolutions.
The star  formation rate is proportional to  the gas amount.  We
confirm that only the  star formation scenarios of elliptical galaxies
are able  to reproduce  the stellar populations  of the  most powerful
radio galaxy  hosts. The robustness of  our model P\'EGASE  is to 
predict  colors  and  SEDs   of  the  typical  observed  templates  at
$\mathrm{z}= 0$.   Statistical values  of such colors  were estimated
after  corrections  (inclination,  extinction,  aperture) by  Fioc  \&
Rocca-Volmerange (1999b). Scenarios of star formation, galactic winds,
infall,  also  respecting  metal  and dust  evolution  constraints  in
P\'EGASE, allow us to predict the details of the evolution of the various
mass components (Fig \ref{figure:Masses}).

Previous studies already tested the validity of the 
scenarios at higher $z$. 
The $U$~to $K$ multi-spectral faint galaxy counts 
from the HDF-N and Hawaii surveys
were fitted with these scenarios (Fioc \& Rocca-Volmerange 1999a).
Another constraint of the validity of the scenarios  
is given by the photometric redshifts computed with the code $Z$-$PEG$.
Used as high $z$\ templates, evolutionary scenarios 
give photometric redshifts close
to  spectroscopic redshifts with $\sigma < $ 0.1
(Le Borgne \& Rocca-Volmerange 2002).

Does  the  observed  dispersion  of the  $K$-$z$~relation  weaken  our
results? One  fundamental property of  the relation is  its uniformity
(Lilly  \&  Longair, 1984)  with  an  increase  of the  dispersion  at
redshifts $z>$ 2 (Eales et al.,  1997; Lacy et al., 2000). McCarthy et
al.  (1999) compared the  present samples  (3CR, 6C  and NIRC)  to the
Molonglo  (MRC/1Jy) sample, complete  in the  sense of  containing $>$
99\% of all sources above its flux limit (S$_{408MHz} >$ 0.95 Jy). The
authors conclude  to the  consistency of the  3CR, MRC and  6C samples
within  1$\sigma$~at $z<0.5$.   At  higher $z$,  Willott et  al.(2003)
combine the 6C, 6C* samples and the 7C sample, complete above the flux
limit  (S$_{151MHz} >$  0.5  Jy) in  three  regions of  the sky.  They
confirm the homogeneity of the stellar populations formed at high $z$~
and following a passive evolution. The variation from $z$=0.05 to 3 is
lower than  5\%, in  agreement with Jarvis  et al. (2001).   They also
found that  the interpretation by  using two models  (no-evolution and
$z_{for}$=5 ) are not in agreement  with the data.  We do not study them in
the  present paper.  Their conclusion  is that  the decrease  of radio
galaxy  masses at  high redshifts  is  small, presuming  a very  rapid
evolution for radio sources at high $z$.

\item Regarding the debate on the IMF as the explanation for divergences
 between observables
and models, we confirm that $K$~magnitudes are faintly sensitive to the IMF (Tinsley, 1972).
However a reasonable value of the lower mass is essential since 
the bulk of stars appearing
in the $K$-band at $z$=0 is made up of low mass stars.
The value of 0.1 $\mathrm{M_\odot}$, adopted for all IMFs, is compatible with
the data and with theoretical star formation principles. By respecting
this constraint on the lower star limit, changing the IMF (including top-heavy) slope will not significantly modify
our conclusions within the error bars.   

\item Remarks about the comparison of dissipative collapse
and hierarchical merging scenarios are still preliminary. However, some 
general ideas may be derived from our conclusions. 
Stellar masses predicted at $z=0$ for ellipticals and spirals 
(respectively 55 $\%$~ and $>$ 45 $\%$~of the initial cloud of mass
$M_{\mathrm{bar}}$) are in 
agreement with typical values for nearby field galaxies. They
range from 10$^{9}\,\mathrm{M_\odot}$~for 
dwarf to ~10$^{12}\,\mathrm{M_\odot}$ for 
giant galaxies. Field galaxies correspond to long time-scales ($\geq$ 10 Gyr)
of less massive clouds, which may reach by hierarchical merging  
a final mass of $10^{12}\,\mathrm{M_\odot}$, the fragmentation limit.
The time-scales of mass accumulation, gas depletion and star formation 
of massive ellipticals at high $z$\ are, however, incompatible with building
blocks of 10$^{5}\,\mathrm{M_\odot}$, proposed by the CDM model.
Our results cannot exclude some merging process of massive 
(a few 10$^{11}\,\mathrm{M_\odot}$) blocks in the dense environment 
of radio galaxies or proto-clusters. 

\item{Is the presence  of black holes in radio  sources linked to host
elliptical  formation?}  To interpret  massive elliptical  galaxies at
high $z$  requires the accumulation  of the dynamical mass on a
short time-scale.   It was suggested  that the embedded black  hole is
associated  with   a  rapid  mass  accumulation   and  star  formation
process. The  more massive  the black hole,  the deeper  the potential
wall;  this   does  not   necessarily  favor  short   time-scales  for
accumulation of  baryonic and stellar  mass in the environment  of the
potential wall.  In the  absence of a  magnetic flux, a  massive cloud
evolves as a  single unit, after having shed  its angular momentum via
instabilities.  The  dense environment may  favor the apparition  of a
black hole (Silk \& Rees, 1998). The high density of galaxy companions
within the ionized  halos of distant radio galaxies  (of critical size
150   kpc)   has   been   shown   by   integral   field   spectroscopy
(Rocca-Volmerange  et  al.,  1994),  and narrow-band  imaging  of  the
Ly$\alpha$~line (Kurk et al.,  2003).  Moreover the interactive r\^ole
of  a  radio  jet  to  trigger  star  formation  is  still  a  debated
subject. Such an hypothesis  is favored by the observational relations
between black  hole masses and absolute blue  luminosities or velocity
dispersion  (Ferrarese  \&  Merritt,   2000;  Gebhardt  et  al.  2000;
Magorrian et al. 1988). However it is too early to identify the nature
of  the relation  between stellar  and  black hole  masses (McLure  \&
Dunlop, 2002).   The bulk  of star formation  could happen  during the
free-fall time while the active  nucleus is fueled. The black hole and
the stellar  mass would then  simultaneously grow; details  of such
processes would, however, need clarification.

Recent deep X-ray surveys which follow the evolution of accretion onto
supermassive black holes  confirm two modes of accretion  and of black
hole growth.  While the rare,  high luminosity QSOs could  form rather
early  in the  universe, a  late evolution  of  low-luminosity Seyfert
populations  is  required by  observations  (Hasinger  et al.,  2003).
Because  high  luminosity  QSOs  are embedded  in  massive  elliptical
galaxies  while  low-luminosity  Seyfert galaxies  are  preferentially
found in spiral galaxies, the two modes of star formation derived from
the $K$-band Hubble diagram can  be related to the two accretion modes
derived  from X-rays.   Moreover  the $K$-band  Hubble diagram  of
sub-mm and  hyperluminous galaxies (Serjeant  et al. ,  2003) recently
confirms that the location of  galaxies on the $K$-$z$~relation may be
related to the presence of the most massive AGNs, in perfect agreement
with our conclusions.
\end{itemize}

A new  factor of uncertainty could be  the ultra-violet continuum
of radio sources.   However because galaxies are observed  in the $K$~
band, only radio galaxies at $z >$ 4 will be affected by this effect.
 
Complementary  constraints  are awaited  from improved  spatial
resolution (morphology, surface brightness, velocity fields).  The $(1
+    z)^4$\ surface   brightness fading    biases   high
$\mathrm{z}$~observations towards  bulges rather than  disks. This may
explain  why,  at  $z  \simeq$  4,  field  galaxies  appear  on  lower
M$_{\mathrm{bar}}$~       sequences       (10$^{9}$~to      10$^{11.5}
\mathrm{M_\odot}$)    than   radio    galaxies    of   high    surface
brightnesses. We  may also evoke  the high density of  galaxies within
their      diameter-limited     sample     to      explain     massive
10$^{12}\mathrm{M_\odot}$\  radio  sources at  high  redshift. Only  a
better spatial resolution, as given by integral field units on the new
generation of  telescopes, will clarify the spatial  confusion of
the central radio galaxy and its galaxy companions.

The baryonic mass predicted by our modeling of luminosity evolution is
a  minimal  value.  From  Fig.  \ref{figure:Masses}  at
$z\simeq$0  (14   Gyr),  a  mass  fraction  $1-M_{gal}$   (35  \%  for
ellipticals, 10\% for  spirals) is still in the  halo. The galaxy halo
could also contain  another baryonic component (such as  hot gas), not
taken into account in our mass budget.

\section{Conclusion}
The main results of this analysis are the following:
 
i) Stellar masses of galaxies, and the corresponding baryonic masses of 
their progenitor clouds,
are estimated from the $\mathrm{L_K}$\ luminosities at any $z$\ 
using robust scenarios of star formation evolution. 
  
ii) The brightest luminosity limit of the Hubble $K$\ diagram 
corresponds to the most massive elliptical galaxies of
baryonic masses M$_{\mathrm{bar,max}} \simeq$~10$^{12}\,\mathrm{M_\odot}$\  
(about 55 \% is in stellar mass at $z$=0). These are the 
typical hosts of powerful
radio galaxies.

iii) The maximum limit of baryonic mass 
M$_{\mathrm{bar,max}}\simeq$~10$^{12}\,\mathrm{M_\odot}$\ ,
estimated from the stellar luminosity in the $K$-band,
corresponds 
to the dynamical parameter M$_{crit}$,  the fragmentation limit 
of the 
cooling-gravitation balance of a self-gravitating cloud.
This 
confirms the validity of the mass estimates from evolution scenarios. 

iv) Radio galaxies of 10$^{12}\,\mathrm{M_\odot}$\ are found at $z > $ 4 with
star formation  and mass accumulation time-scales shorter than 1Gyr. 
The radius of the ionized cocoon is about $r_{crit}$=75kpc.
Such extreme
time-scales in powerful radio galaxies suggest that the formation 
of a massive black hole 
embedded in the host galaxy is not independent of the
star formation process.

v) The choice of the IMF slope is not crucial but the lower star mass
needs to be $\simeq$ 0.1 to form the bulk of low mass stars 
observed in templates of elliptical galaxies at $z$=0.

vi) Galaxy deep surveys are covered by sequences  
of lower baryonic masses from  
M$_{\mathrm{bar}}$= ~10$^{9}$\ to 10$^{11}\,\mathrm{M_\odot}$
in which the evolution by hierarchical merging, if it is proven, 
remains however mass limited by fragmentation.

For the future, we will need to apply such a method of mass determination
to the galaxy populations of large surveys, in particular discovered 
either by
SIRTF or by the ground-based 10m telescopes in the stellar infra-red
to follow the evolution processes of galaxies. \\ 

\begin{acknowledgements}
We  would like to warmly thank the referee Carol Lonsdale 
for her constructive and detailed
comments. European Community programmes (Marie-Curie fellowship for CdB and 
TMR POE fellowship for EM) helped significantly to achieve  the present work. 
\end{acknowledgements}

\end{document}